# Comment on "Spinless impurities in High-$T_c$ Cuprates: Kondo-like behavior"

In a recent Letter [1] Bobroff *et al.* compare the effects of Zn- and Li-substitution in $YBa_2Cu_3O_{7-\delta}$ (Y-123) and $Y_{0.8}Ca_{0.2}Ba_2Cu_3O_{7-\delta}$ using $^{89}Y$ and $^7Li$ NMR. $^7Li$ NMR allows precise determination of the T-dependence of the local spin susceptibility especially in the overdoped region where the $^{89}Y$ satellites (which also probe the local susceptibility) could no longer be distinguished from the main resonance (probing the bulk susceptibility). From the T-dependence of the local susceptibility, which progressed from Curie-like to a $1/(T+\theta)$ Kondo-like variation, these spinless substitutions were both inferred to induce a local moment on the nearest-neighbour (n-n) Cu sites and, allowing for distribution of Li over both Cu1 and Cu2 sites, the authors found Zn and Li to equally suppress $T_c$. These conclusions would appear to be sound. They then draw two further conclusions which are less secure. Firstly, that the strong impurity scattering is associated with the absence of a spin and secondly, that even in overdoped materials there is still a substantial induced moment.

The fact that Ni is an equally strong scatterer, reducing $T_c$ in $YBa_2Cu_4O_8$ (Y-124) and $La_{2-x}Sr_xCuO_4$ just as rapidly as Zn [2] would appear to counter the first conclusion. The well-known weaker suppression in Y-123 is a result of distribution between Cu1 and Cu2 sites and can be enhanced, for example, by phase conversion to Y-123 from Zn-substituted Y-124 in which only the Cu2 sites are occupied [3].

Turning to the second issue, the local AF background is considered to leave an unbalanced moment on the four n-n Cu sites surrounding the spin vacancy. Elsewhere [4], we have shown that these short-range AF correlations appear to die out rather suddenly at the critical doping point, $p_{crit} \approx 0.19$, where the pseudogap disappears. Indeed Mendels *et al.* [5] find that the Zn-induced moment falls rapidly with increasing doping in Y-123 towards zero at full oxygenation, i.e. at $p_{crit}$ [4]. A Li-induced moment in the heavily overdoped region would appear to contradict this scenario. However, we show here that a Kondo-like overdoped T-dependence does not imply the existence of a Li-induced moment but is *intrinsic*. Fig. 1(a) shows $^{89}K_s$ for a succession of overdoped samples of *pure* $Y_{0.8}Ca_{0.2}Ba_2Cu_3O_{7-\delta}$. The suppression of $K_s$ in the underdoped and optimally doped samples is associated with the pseudogap [6,4] and modelled in the figure by a triangular gap [4]. For $p>p_{crit}$, $^{89}K_s$ exhibits an intrinsic upturn that is well fitted by a $1/(T+\theta)$ dependence (dashed curves). This is almost certainly the origin of the Kondo-like susceptibility probed by $^7K_s$ and is unrelated to impurities. Thus, in the overdoped region the upturn in $^7K_s(T)$ reflects the intrinsic bulk susceptibility not the extrinsic local susceptibility due to an induced moment. We also show in Fig. 1(b) the T-dependence of $S^{el}/T$, where $S^{el}$ is the electronic entropy, for overdoped $Bi_2Sr_2CaCu_2O_{8+\delta}$ [7]. This also is well fitted by the same $1/(T+\theta)$ dependence reflecting an intrinsic low-energy peak in the density of states (DOS) which grows with overdoping and is the cause of the upturn in both the entropy and susceptibility (which are linearly related via the Wilson ratio [8]). This peak in the overdoped DOS is also evident in symmetrised ARPES spectra [9]. It would seem, then, that there is no induced moment in the overdoped region and this is the reason why the $^{89}Y$ n-n satellite coincides with the main resonance there.

J.L. Tallon, J.W. Loram* and G.V.M. Williams
Industrial Research, PO Box 31310, Lower Hutt, N.Z.
*Cambridge University, Cambridge, CB3 0HE, U.K.

[1] J. Bobroff *et al.*, Phys. Rev. Lett. **83**, 4381 (1999).
[2] G.V.M. Williams *et al.*, Phys. Rev. B **61**, 4319 (2000).
[3] G.V.M. Williams and S. Krämer (to be published).
[4] J.L. Tallon and J.W. Loram, cond-mat/0005063
[5] P. Mendels *et al.*, Europhys. Lett. **46**, 678 (1999).
[6] H. Alloul *et al.*, Phys. Rev. Lett. **63**, 1700 (1989).
[7] J.W. Loram *et al.*, Physica C (in press).
[8] J.W. Loram *et al.*, Physica C **235-240**, 134 (1994).
[9] M.R. Norman *et al.*, Phys. Rev. B **57**, R11093 (1998).

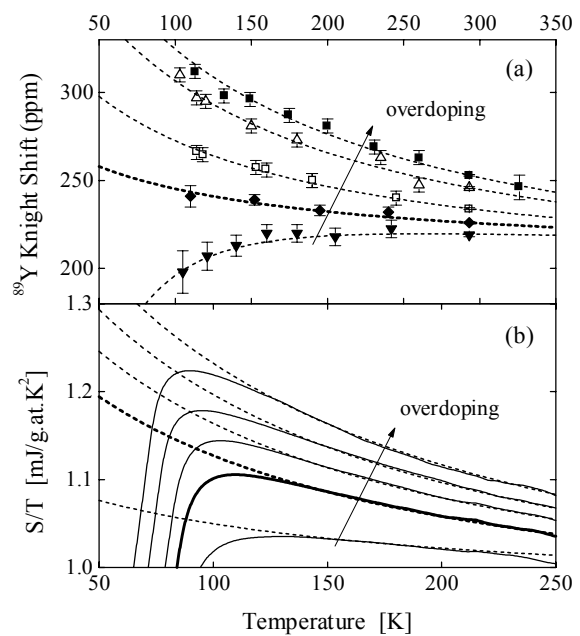

Fig. 1. $^{89}Y$ NMR Knight shift, $^{89}K_s$, for overdoped $Y_{0.8}Ca_{0.2}Ba_2Cu_3O_{7-\delta}$ (a) and electronic entropy $S/T$ for overdoped Bi-2212 (b) fitted to $1/(T+\theta)$ with $\theta=110K$ (dashed curves). Bold curves denote critical doping, $p_{crit}=0.19$.